\documentclass[11pt]{article}
\usepackage{graphicx}
 
\begin{document}

\title{Finite temperature dynamic susceptibility of the free Bose
gas}

\author{F.Mazzanti$^{1,2}$ and A.Polls$^3$  \\  \\
        $^1$ {\small Institut f\"ur Theoretische Physik, 
	             Johannes Kepler Universit\"at Linz,}
	     {\small A-4040 Linz, Austria} \\
        $^2$ {\small Departament d'electr\`onica,
	             Enginyeria i Arquitectura La Salle,} \\
	     {\small Pg. Bonanova 8,  Universitat Ramon Llull,
	             E-08022 Barcelona, Spain} \\
        $^3$ {\small Departament d'Estructura i Constituents
                     de la Mat\`eria,} \\
             {\small Diagonal 645, Universitat de Barcelona,
                     E-08028 Barcelona, Spain} }
%        $^4$ {\small School of Physics and Astronomy, 
%                     University of Minnesota,} \\
%             {\small 116 Church Street S.E., Minneapolis, 
%                     Minnesota 55455} }

\maketitle

\begin{abstract}
A detailed calculation of the real part of the finite temperature
dynamic susceptibility of the free Bose gas is presented. After a
short discussion on the different ways in which it can be calculated
for temperatures above and below the Bose--Einstein transition
temperature, its main properties and its evolution with $q$ and $T$
are analyzed. Finally, expressions for the lowest order energy
weighted sum rules are also derived and studied.
\end{abstract}

\maketitle

\pagebreak

The clear resolution of Bose--Einstein condensates in clouds of Alkali
atoms at ultralow temperatures has revived the interest in the study
of free and weakly interacting, homogeneous and inhomogeneous Bose
systems during the last years~\cite{dalf}. Recent experimental
evidence~\cite{exp1,exp2} indicates that Bragg scattering can be
succesfully applied to measure the dynamic structure function
$S(q,\omega)$ of trapped Bose--Einstein condensates, and hence to
directly probe their momentum distribution~\cite{strin}.

During the last thirty years, much experimental and theoretical
efforts have also been devoted towards the resolution of the
condensate from neutron scattering in other, more correlated systems
like pure $^4$He and $^3$He--$^4$He mixtures.  However, none of the
available formalisms have been able to accurately describe the
influence of the temperature on the response, and particularly its
influence on the condensate fraction, both at low and high momentum
transfer \cite{gly1}.
 
In dilute systems, like the Bose--Einstein condensates of trapped
alkali atoms, correlations are so weak that the response is expected
to be qualitatively well described, at least at low momentum transfer,
starting from the response of a free system and introducing the effect
of the correlations through the Random Phase Approximation (RPA). In a
previous work~\cite{sqwTbos}, a detailed description of the finite
temperature dynamic structure function of the free Bose gas has been
presented and discussed, focusing particularly on how $q$ and $T$
affect the total response and its coherent and incoherent parts. In
this work we extend the analysis to the real part of the dynamic
susceptibility $\chi(q,\omega;T)$, which is the building block from
which the RPA is constructed.

The dynamic susceptibility of an homogeneous free system is given
by~\cite{fetwal,pin1} 
\begin{eqnarray}
\chi(q,\omega;T) & = & 
{1\over N} \sum_{\bf p} n_{\bf p} \left[
{1\over \omega - t_{{\bf p}+{\bf q}} + t_p + i\eta} -
{1\over \omega - t_p + t_{{\bf p}-{\bf q}} + i\eta} 
\right]
\nonumber \\ [2mm] 
& = & 
{\nu \over (2\pi)^3\rho} \int d{\bf p}\, n(p) \left[
{1\over \omega - t_{{\bf p}+{\bf q}} + t_p + i\eta} -
{1\over \omega - t_p + t_{{\bf p}-{\bf q}} + i\eta} 
\right] \ ,
\label{c1}
\end{eqnarray}
where $\rho$ is the density, $t_p=p^2/2m$ is the free kinetic energy
spectrum, $n(p)$ is the momentum distribution proportional to the
occupation probability of each single--particle state of definite
momentum, and $\nu$ is the spin-isospin degeneracy which will be taken
equal to unity throughout this letter.

In the specific case of a free Bose gas case, $n(p)$ is the sum of a
condensate and a non--condensate terms
\begin{equation}
n(p) = (2\pi)^3\rho\,n_0(T)\delta({\bf p})  + 
{1\over z^{-1} e^{\beta p^2/2m} - 1} \ ,
\label{c2}
\end{equation}
where $\beta=1/T$ is the inverse of the temperature and $z$ is the
fugacity, related to the chemical potential $\mu$ and the
temperature through $z=e^{\beta\mu}$. The condensate fraction value
$n_0(T)$ varies with the temperature according to a $T^{3/2}$ law
\begin{equation}
n_0(T) = 1 - \left( {T \over T_c} \right)^{3/2} \ ,
\label{c2c}
\end{equation}
where $T_c\approx 3.31 \rho^{2/3}/m$ is the temperature at which the
Bose--Einstein condensation sets in. Finally, the chemical potential
$\mu(T)$ is fixed at each $T$ by imposing the particle normalization
condition
\begin{equation}
{1\over (2\pi)^3\rho} \int d{\bf p} \,n(p) = 1 \ .
\label{c2b}
\end{equation}

At finite temperature, the imaginary part of the dynamic
susceptibility is related to the dynamic structure function $S(q,
\omega; T)$ through the general relation~\cite{pin1}
\begin{equation}
{\rm Im} \left[ \chi(q,\omega;T)\right] = -\pi 
\left[ S(q,\omega;T) - S(q,-\omega;T) \right] \equiv
-\pi \left(1-e^{-\beta\omega} \right)
S(q,\omega;T) \ ,
\label{c3}
\end{equation}
where use has been made of the detailed balance condition, which
relates the $\omega>0$ and the $\omega<0$ contributions to the finite
temperature response in the form
\begin{equation}
S(q,-\omega; T) = e^{-\beta\omega} S(q,\omega; T) \ .
\label{c3b}
\end{equation}

The finite temperature $S(q, \omega; T)$ of the free Bose gas has been
calculated in~\cite{sqwTbos} and yields, in terms of a new set of
dimensionless variables that will be used throughout this work $\tilde
q=q/k_B$, $\tilde\omega=\omega/\epsilon_B$ and $\tilde T=T/\epsilon_B$
where $\epsilon_B=k_B^2/2m$ and $k_B=\rho^{1/3}$ define the energy and
momentum scales,
\begin{eqnarray}
\tilde S(\tilde q, \tilde\omega; \tilde T) & = & 
{n_0(\tilde T)\over 1-e^{-\tilde\omega/\tilde T}} 
\left[ \delta\!\left(\tilde\omega - \tilde q^2\right) - 
\delta\!\left(\tilde\omega + \tilde q^2\right) \right]
\nonumber \\ [2mm]
& & -{\tilde T\over 16\pi^2 \tilde q}
{1\over \left(1-e^{-\tilde\omega/\tilde T} \right)} 
\ln\left[ 
{ 1 - z e^{-(\tilde\omega/\tilde q - \tilde q)^2/4\tilde T}  \over 
  1 - z e^{-(\tilde\omega/\tilde q + \tilde q)^2/4\tilde T}} \right] \ .
\label{d1}
\end{eqnarray}

The first term in this expression takes into account the contribution
of the atoms in the condensate and appears in the form of two delta
peaks centered at the quasielastic recoil energies
$\tilde\omega=\pm\tilde q^2$. The second term describes the
contribution coming from the rest of the atoms and shows itself as a
function with two well defined peaks of finite width and height
centered at the same energies.

The real part of the dynamic susceptibility splits also in two pieces,
corresponding to the condensate and non--condensate contributions,
respectively. The condensate part $Re[\tilde\chi_c(\tilde q,
\tilde\omega; \tilde T)]$ is equal to $n_0(T)$ times the real part of
the $\tilde T=0$ dynamic susceptibility
\begin{equation}
Re[\tilde\chi_c(\tilde q, \tilde\omega; \tilde T)] = 
n_0(T) Re \left[
{1 \over \omega - \tilde q^2 + i\eta} - 
{1 \over \omega + \tilde q^2 + i\eta} \right] \ ,
\label{d1b}
\end{equation}
thus diverging at $\tilde\omega=\pm\tilde q^2$.  On the other hand,
the non--condensate part is a genuine contribution due to the finite
temperature and is the main quantity analyzed in this work.

The real part of the dynamic susceptibility can be calculated in different
ways, depending on whether the temperature is above or below the
Bose--Einstein transition temperature. In a standard treatment, one can
just evaluate it by using a Kramers--Kronig relation
\begin{equation}
Re\left[\tilde\chi(\tilde q, \tilde\omega;\tilde T) \right] = 
{1\over \pi} {\mathcal P} \int d\tilde\omega' \,
{Im\left[\tilde\chi(\tilde q, \tilde\omega; \tilde T)\right] \over 
\tilde\omega' - \tilde\omega} \ ,
\label{d2}
\end{equation}
where ${\mathcal P}$ stands for Principal Value Integration. This
expression relies on the implicit assumption that
$Im\left[\tilde\chi(\tilde q, \tilde\omega; \tilde T)\right]$ is an
analytic function in the upper half of the complex plane, and that it
decreases fast enough with increasing energy. Notice also that
Eqs.~(\ref{c3}) and ~(\ref{d2}) imply that $Im[\tilde\chi(\tilde q,
\tilde\omega; \tilde T)]$ and $Re[\tilde\chi(\tilde q, \tilde\omega;
\tilde T)]$ are odd and even functions of $\omega$, respectively.

Alternatively, one can evaluate $Re\left[\tilde\chi(\tilde q,
\tilde\omega; \tilde T)\right]$ as an infinite series, generalizing
the result originally derived by Khanna and Glyde~\cite{khangly} for
the free Fermi gas. Two different expressions are obtained in this
way. In both cases, the dynamic susceptibility is written in the form
\begin{equation}
\tilde\chi(\tilde q, \tilde\omega; \tilde T) = 
\tilde\chi(\tilde q, 0; \tilde T) + 
\int_0^{\tilde\omega} d\xi {\partial \over \partial\xi} 
\tilde\chi(\tilde q, \xi; \tilde T)
\label{d3}
\end{equation}
where the partial derivative in the rhs can be evaluated from the
definition in Eq.~(\ref{c1}) and is expressed as an integral that may
be carried out on the complex plane. As a result one gets an infinite
series corresponding to a sum of residues, that has to be inserted in
Eq.~(\ref{d3}) and integrated to get the total dynamic susceptibility.

In a first approach, one integrates this series term by term to end up
with 
\begin{eqnarray}
Re\left[\tilde\chi(\tilde q, \tilde\omega; \tilde T)\right] 
& = & 
Re\left[\tilde\chi(\tilde q, 0; \tilde T)\right]  
\nonumber \\ 
& + & {\tilde T \over 8\pi\tilde q} \sum_n \left[ 
\tan^{-1}\left( {2\tilde y_1\tilde b_n \over \tilde y_1^2 - (\tilde a_n^2
+ \tilde b_n^2)} \right) - \tan^{-1}\left( {2 \tilde y_2 \tilde b_n
\over \tilde y_2^2 - (\tilde a_n^2 + \tilde b_n^2)} \right) \right] \ ,
\label{e1}
\end{eqnarray}
where the index $n$ runs from $0$ to $\infty$.  In
Eq.~(\ref{e1}), $\tilde a_n$ and $\tilde b_n$ are the real and
imaginary parts of the poles of the momentum distribution laying in
the first quadrant
\begin{eqnarray}
\tilde a_n & = & 
{1\over\sqrt{2}} \left[ \tilde\mu + \left( \tilde\mu^2 + 4\pi^2 n^2
\tilde T^2 \right)^{1/2} \right]^{1/2}
\nonumber \\
\tilde b_n & = & 
{1\over\sqrt{2}} \left[ -\tilde\mu + \left( \tilde\mu^2 + 4\pi^2 n^2
\tilde T^2 \right)^{1/2} \right]^{1/2} \ ,
\label{e2}
\end{eqnarray}
while $\tilde y_1$ and $\tilde y_2$ are the usual West scaling
variables, related to the momentum and the energy transfer through
\begin{equation}
\tilde y_1 = {1\over 2}
\left({\tilde\omega \over \tilde q} - \tilde{q} \right)
\,\,\,\,\,\,\,\, \& \,\,\,\,\,\,\,\, 
\tilde y_2 = {1\over 2}
\left({\tilde\omega \over \tilde q} + \tilde{q} \right) \ .
\label{e3}
\end{equation}

Alternatively, one can first evaluate the sum of all the residues
contributing to the energy derivative of the real part of the
susceptibility 
\begin{eqnarray}
{\partial \over \partial\tilde\omega} 
Re\left[\tilde\chi(\tilde q, \tilde\omega; \tilde T)\right] & = & 
{\tilde T \over 8\pi\tilde q^2 } \sum_{n>0} \Bigg\{ 
2\pi n\tilde T \left( {\tilde a_n\over \tilde a_n^2+\tilde b_n^2}
\right) \left[ 
{\tilde y_2^2 \over \left(\tilde y_2^2 - \tilde\mu \right)^2 + 
4 \pi^2 \tilde T^2 n^2 } - 
{\tilde y_1^2 \over \left(\tilde y_1^2 - \tilde\mu \right)^2 + 
4 \pi^2 \tilde T^2 n^2 } \right]
\nonumber \\ 
& & - \left({\tilde b_n \over \tilde a_n^2 + \tilde b_n^2}\right) 
\left[ {\tilde y_2^2 \left(\tilde y_2^2 - \tilde\mu \right) \over 
\left( \tilde y_2^2 - \tilde\mu \right)^2 + 4\pi^2\tilde T^2 n^2 } - 
{\tilde y_1^2 \left(\tilde y_1^2 - \tilde\mu \right) \over 
\left( \tilde y_1^2 - \tilde\mu \right)^2 + 4\pi^2\tilde T^2 n^2 }
\right] \Bigg\} \ .
\label{e5}
\end{eqnarray}
and afterwards perform the integration. Although this method seems to
be more elaborated than the simple sum in Eq.~(\ref{e1}), it is
usually preferred since the former expression is known to converge
rather slowly. In both cases, however, $Re\left[\tilde\chi(\tilde q,
0; \tilde T)\right]$ is a required quantity that can be obtained,
proceeding as before, in the form of a series
\begin{eqnarray}
Re\left[\tilde\chi(\tilde q, 0; \tilde T)\right] & = &
Re\left[\tilde\chi(0, 0; \tilde T)\right] 
\nonumber \\ 
& & - {\tilde T\over 32\pi\tilde q} \int_0^{\tilde q} d\tilde k\, 
\tilde k^2 \sum_n 
{\tilde a_n 2\pi n\tilde T - \tilde b_n \left( {\tilde k^2\over 4} -
\tilde\mu \right) \over 
\left( \tilde a_n^2 + \tilde b_n^2 \right) \left( 
\left( {\tilde k^2 \over 4} - \tilde\mu \right)^2 + 
4\pi^2 \tilde T^2 n^2 \right) }
\label{e6}
\end{eqnarray}
where
\begin{equation}
Re\left[\tilde\chi(0,0; \tilde T)\right] = 
-{1\over 2\pi^2} \int_0^\infty d\tilde p \,n(\tilde p) \ .
\label{e7}
\end{equation}
and the sum extending from $0$ to $\infty$ with $\tilde a_n$ and
$\tilde b_n$ given in Eq.~(\ref{e2}).

Finally, $Re[\tilde\chi(\tilde q, \tilde\omega; \tilde T)]$ can also
be evaluated by direct numerical integration of Eq.~(\ref{c1}). After
simple manipulations, one finds
\begin{eqnarray}
Re[\tilde\chi(\tilde q, \tilde\omega; \tilde T)] & = &
{1\over 8\pi\tilde q} Re \int d\tilde p\, \tilde p n(\tilde p)
\left[ \ln\left({\tilde y_2 - \tilde p + i\epsilon
\over \tilde y_2 + \tilde p + i\epsilon}\right) - 
\ln\left({\tilde y_1 - \tilde p + i\epsilon \over \tilde y_1 + \tilde p +
i\epsilon} \right) \right]
\nonumber \\ 
& = & {1\over 8\pi\tilde q} 
{\mathcal P} \int_0^\infty d\tilde p\, \tilde p 
n(\tilde p) \left[ 
\ln\left|{\tilde y_2 - \tilde p \over \tilde y_2 + \tilde p} \right| - 
\ln\left|{\tilde y_1 - \tilde p \over \tilde y_1 + \tilde p} \right|
\right] 
\label{f1}
\end{eqnarray}
which can be safely carried out since their singularities are of the
integrable type.

For temperatures above $\tilde T_c$, all three methods are equivalent
and any of them can be used to obtain the non--condensate contribution
to the real part of the dynamic susceptibility. Below $\tilde T_c$,
however, only the direct integration method and the Kramers--Kronig
relation can be applied, even though in the latter case the
analyticity condition required for $Im[\tilde\chi(\tilde q,
\tilde\omega; \tilde T)]$ is clearly violated since at those
temperatures the chemical potential vanishes and the fugacity $z$ goes
to $1$, therefore the function presents two singularities on the real
axis located at the quasielastic energies $\tilde\omega=\pm\tilde
q^2$. However, these are logarithmic singularities that can be
integrated on the complex plane and yield no additional contribution
to the integrals. On the other hand, none of the series methods
mentioned above can be used because below $\tilde T_c$ the momentum
distribution grows as $\tilde p^{-2}$ for small $\tilde p$, thus
making $Re[\tilde\chi(0,0;\tilde T)]$ in Eq.~(\ref{e7}) diverge.

Since $\tilde\mu(\tilde T\leq\tilde T_c)=0$, the temperature
dependence of $Re[\tilde\chi(\tilde q, \tilde\omega; \tilde
T\leq\tilde T_c)]$ can be easily extracted once it is written in terms
of two new variables $Q=\tilde q/\sqrt{\tilde T}$ and
$\nu=\tilde\omega/\tilde T$. Moreover, one can define a new
function 
\begin{equation}
Re[\hat\chi(Q,\nu)] \equiv {1\over\sqrt{\tilde T}} Re\left[\chi(\tilde q,
\tilde\omega; \tilde T)\right]
\label{f2}
\end{equation}
that does not depend on the temperature when $Q$ and $\nu$ are taken
as the new independent variables. Actually, this scaling property is
also satisfied by the imaginary part of $\tilde\chi(\tilde q,
\tilde\omega, \tilde T)$ which inherits it from the $\tilde
T\leq\tilde T_c$ behaviour of the dynamic structure function of the
free Bose gas~\cite{sqwTbos}.

\begin{figure}
\begin{center}
\includegraphics[height=11cm]{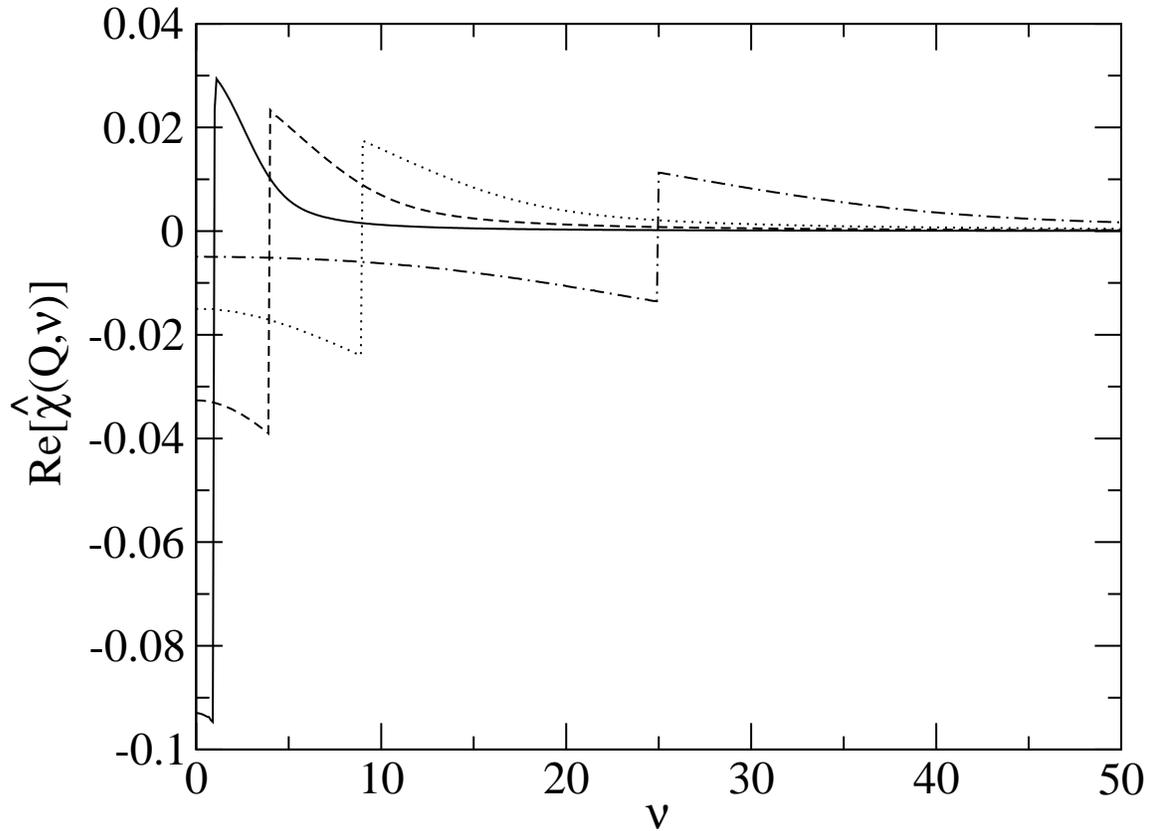}
\caption{Real part of $\hat\chi(Q,\nu)$ at positive $\nu$ and for
several values of $Q$. The solid, dashed, dotted and dash-dotted lines
correspond to Q=1, 2, 3 and 5 respectively}
\label{fig-1}
\end{center}
\end{figure}

$Re[\hat\chi(Q,\nu)]$ is shown in Fig.~(\ref{fig-1}) for several
values of $Q$. At low $\nu$, $Re[\hat\chi(Q,\nu)]$ is negative and
discontinuous at $\nu=Q^2$, jumping to positive values and afterwards
decaying to 0. The discontinuity at $\nu=Q^2$ is a direct consequence
of the divergent behaviour of the momentum distribution at low
momenta, and is therefore directly related to the presence of a Bose
condensate. On the other hand, the decay of $Re[\hat\chi(Q,\nu)]$ at
large energies is of the form $(1-n_0(T))2Q^2/\nu^2 \tilde T^{3/2}
\equiv 2Q^2/\nu^2 \tilde T_c^{3/2}$. This last property is mirrored
from the fact that the real part of the dynamic susceptibility is
related to the dynamic structure function through Eqs.~(\ref{d2})
and~(\ref{c3}), and this together with the $f$--sum rule satisfied by
$\tilde S(\tilde q, \tilde\omega; \tilde T)$ is enough to determine
the analytical behaviour of the real part of the susceptibility at
large energies~\cite{pin1}. Finally, the factor $1-n_0(T)$ reflects
the fact that we are only looking at the non--condensate contribution
to the susceptibility, while the condensate term adds the extra
strength required to have a total decay of the form $2\tilde
q^2/\tilde\omega^2$ as expected.

With rising momentum transfer the discontinuity shifts to higher
$Q$'s, the strength at low energies and around $Q^2$ is depressed, but
the energy range where $Re[\hat\chi(Q,\nu)]$ presents a significant
contribution is increased. Due to the scaling in $Q$ and $\nu$, the
effect of fixing $\tilde T$ and rising $\tilde q$ is equivalent to the
effect of fixing $\tilde q$ and decreasing $\tilde T$. Therefore, the
evolution with $\tilde T$ of $Re[\tilde\chi(\tilde q, \tilde\omega;
\tilde T)]$ at fixed $\tilde q$ can be read in the figure by moving
from higher to lower $Q$'s, while the evolution with $\tilde q$ at
fixed $\tilde T$ is represented by the same sequence of curves but in
reverse order. At $\tilde T=0$ the non--condensate contribution to
the dynamic susceptibility vanishes and only the condensate term is
left.

\begin{figure}
\begin{center}
\includegraphics[height=11cm]{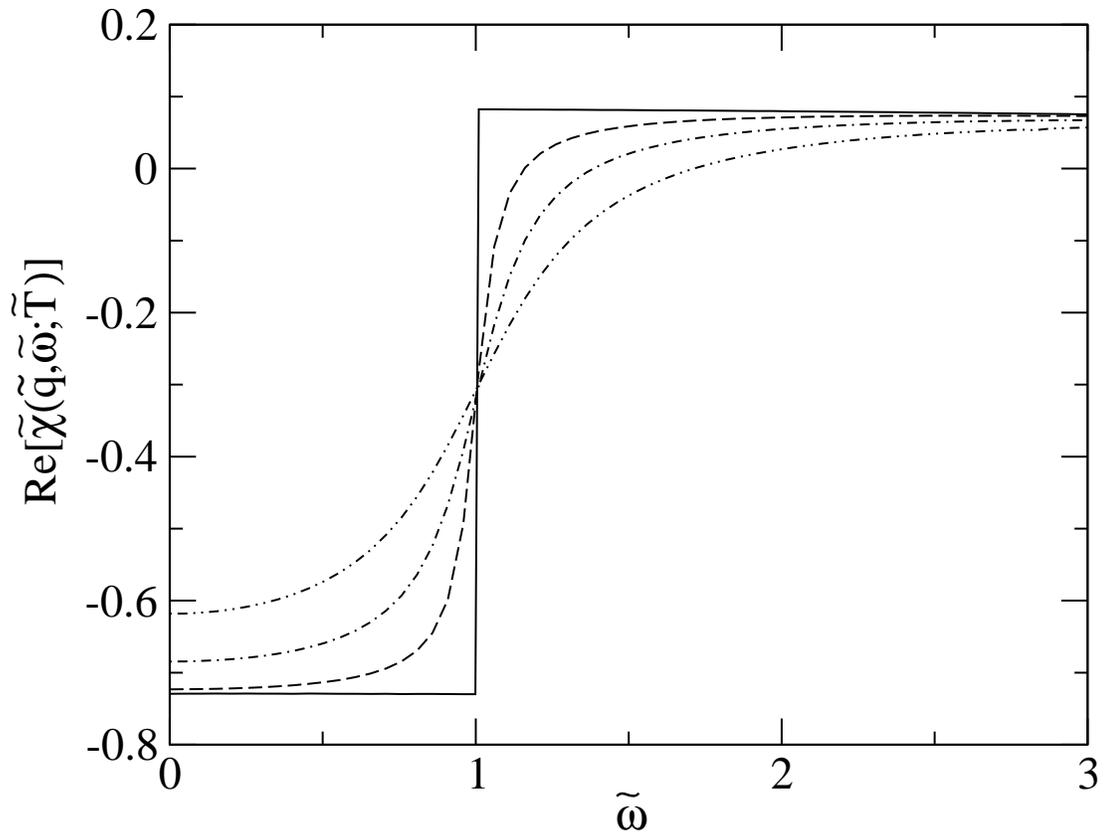}
\caption{ Real part of $\tilde\chi(\tilde q,\tilde\omega; \tilde T)$
at temperatures slightly below and above the Bose--Einstein transition
temperature $\tilde T_c \approx 6.62$.  Solid line: $\tilde T=6.5$,
dashed line: $\tilde T=6.7$, dot--dashed line: $\tilde T=6.8$, and
dot--dot--dashed line: $\tilde T=7$.}
\label{fig-2}
\end{center}
\end{figure}

When $\tilde T$ crosses the Bose--Einstein transition temperature, the
discontinuity at $\tilde\omega=\tilde q^2$ is smeared out and
$Re[\tilde\chi(\tilde q, \tilde\omega; \tilde T)]$ becomes a
continuous function of $\tilde\omega$, as shown in Fig.~(\ref{fig-2})
for a momentum transfer of $\tilde q=1$. This is due to the fact that,
above $\tilde T_c$, the chemical potential is no longer $0$ and thus
the $\tilde p \to 0$ divergence in the momentum distribution is
removed.  Actually the process is quite fast, and already at $\tilde
T=7$ and a momentum transfer of $\tilde q=1$, no apparent trace of a
discontinuity in the real part of the dynamic susceptibility is left.

\begin{figure}
\begin{center}
\includegraphics[height=19cm]{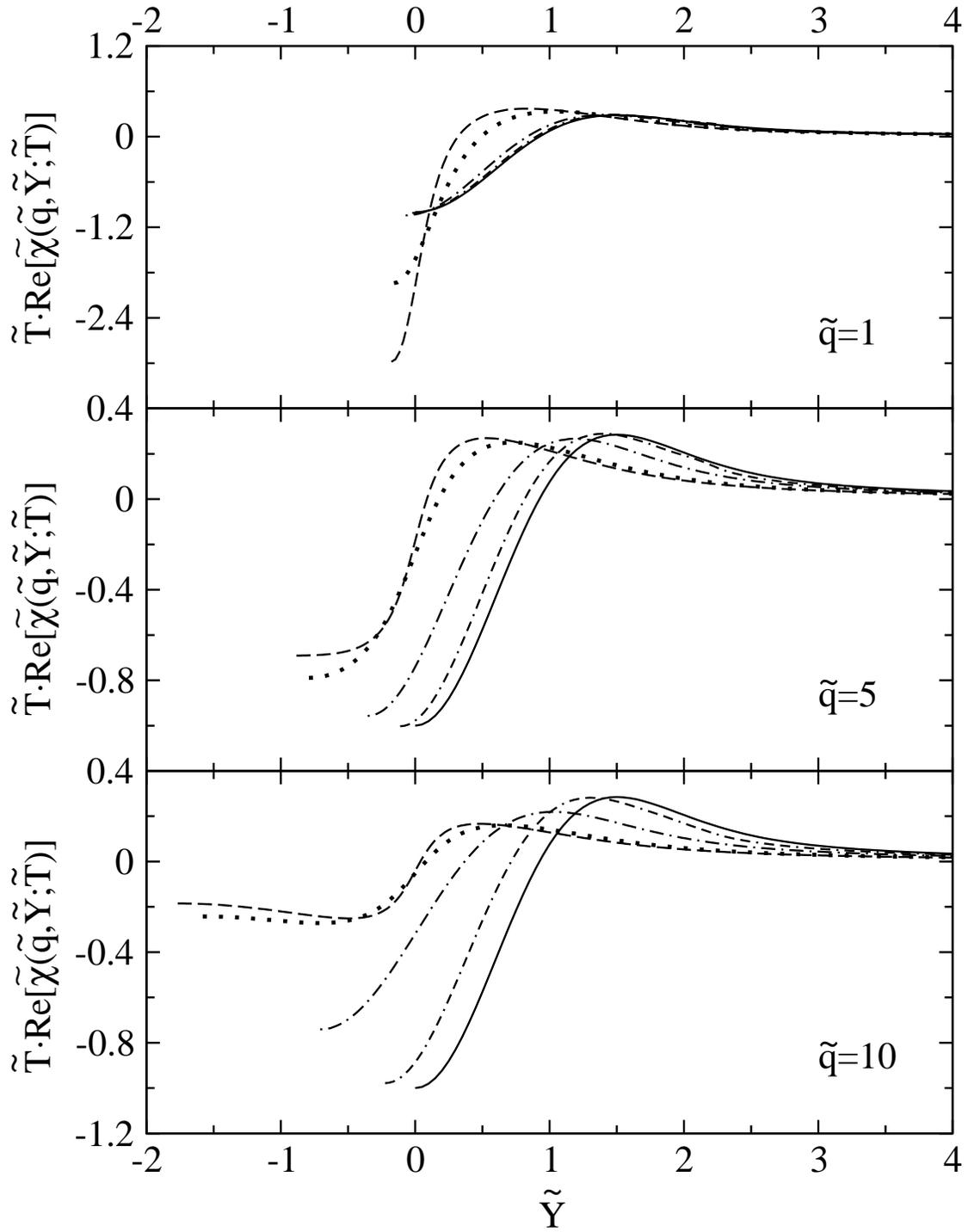}
\caption{Evolution of $\tilde T Re[\tilde\chi(\tilde q, \tilde Y;
\tilde T)]$ with $\tilde T$ and for three values of the momentum
transfer $\tilde q$. The dashed, dotted, dot--dashed and
dot--dashed--dashed lines stand for $\tilde T= 8, 10,50$ and $500$
respectively, while the solid line corresponds to the classical limit}
\label{fig-3}
\end{center}
\end{figure}

The evolution with $\tilde T$ of the non--condensate contribution to
the real part of the dynamic susceptibility for $\tilde T > \tilde
T_c$ and for three values of the momentum transfer is sketched in
Fig.~(\ref{fig-3}).  The actual function depicted is $\tilde T
Re[\tilde\chi(\tilde q, \tilde Y; \tilde T)]$ with $\tilde
Y=(\nu/Q-Q)/2$, since in the high temperature limit the susceptibility
of the free Bose gas approaches the $\tilde T\to\infty$ classical
prediction which for energies around the quasielastic peak
$\tilde\omega=\tilde q^2$ can be casted in the form
\begin{equation}
\tilde T Re[\tilde\chi_{cl}(\tilde q, \tilde Y; \tilde T\to\infty)] = 
-{1\over \sqrt{\pi}} {\mathcal P} \int_{-\infty}^\infty 
dz\, {z e^{-z^2} \over z - \tilde Y}  \ ,
\label{g1}
\end{equation}
thus being a function of $\tilde Y$ alone.  Since
$Re[\tilde\chi(\tilde q, \tilde\omega; \tilde T)]$ is and even
function of $\tilde\omega$, only the contribution at positive energies
is shown, so the initial point in each curve is $\tilde Y=-Q/2$. As it
can be seen from the figure, the departure from the classical limit at
low temperatures is sizeable, while the Bose prediction gets closer to
the classical one when the temperature is risen, as expected. However,
the way in which the classical limit is reached depends on the
momentum transfer. At low $\tilde q$, the low temperature
susceptibility rapidly varies from negative values below the classical
prediction to positive ones above it. With increasing momentum
transfer, the strength of the Bose function at fixed temperature is
progressively reduced, but the range of maximum variation in which the
function goes from negative to positive values grows. In all cases,
however, the large $\tilde Y$ tails of the Bose function conform to
the classical prediction at any temperature, a fact that should not
surprise since as commented above the dynamic susceptibility is known
to decrease as $2\tilde q^2/\tilde\omega^2$ independently of the
temperature and the statistics.

\begin{figure}
\begin{center}
\includegraphics[height=15cm]{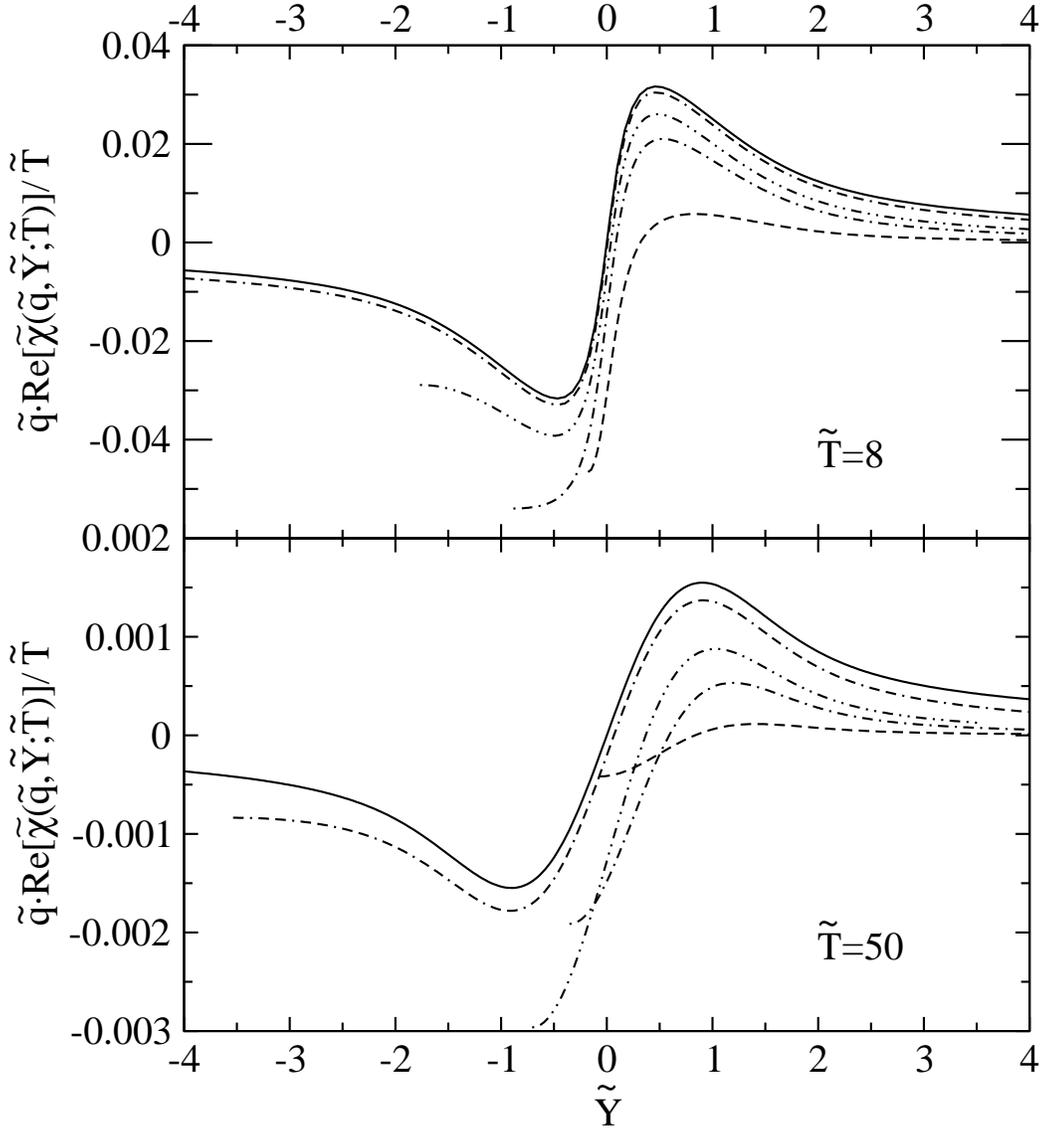}
\caption{Momentum dependence of the $\tilde q Re[\tilde\chi(\tilde q,
\tilde Y; \tilde T)]/\tilde T$ at two different temperatures. The
dashed, dot--dashed, dot--dot--dashed and dash-dash-dotted lines stand
for $\tilde q=1, 5, 10$ and $50$ respectively, while the solid line
corresponds to the limit $\tilde q \rightarrow \infty$.}
\label{fig-4}
\end{center}
\end{figure}

The behaviour of the real part of the dynamic susceptibility at fixed
$\tilde T$ larger than $\tilde T_c$ and for a sequence of increasing
values of the momentum transfer $\tilde q$ is depicted in
Fig.~(\ref{fig-4}). In this case, the $\tilde q \to \infty$ limit of
$Re[\tilde\chi(\tilde q, \tilde Y; \tilde T)]$ at finite $\tilde Y$
can be obtained from the leading term in a $1/\tilde q$ expansion, a
property that is inherited from the high $\tilde q$ behaviour of the
dynamic structure function~\cite{gerosmi}. As in the latter case and
for the free Bose gas~\cite{sqwTbos}, the function
\begin{equation}
{\tilde q \over \tilde T}Re[\tilde\chi(\tilde q, \tilde Y; \tilde T)]
\label{g2}
\end{equation}
approaches a fixed curve when $\tilde q \to \infty$ that depends on
the temperature only through the chemical potential. This limiting
function is represented with a solid line in the figure, while the
other curves show the way in which this limit is approached with
increasing $\tilde q$. Notice that, contrary to what happened in the
previous case, increasing $\tilde q$ at fixed $\tilde T$ makes the
initial point $\tilde Y=-Q/2$ in the limiting curve go to $-\infty$,
thus emphasizing that the region of maximal variation of
$Re[\tilde\chi(\tilde q, \tilde Y; \tilde T)]$ lies always around
$\tilde Y\approx 0$ corresponding to energies close to the quasielastic
peak $\omega=\tilde q^2$. Notice also that the limiting curve is
always reached from the {\em right}, as 
$(\tilde q/\tilde T) Re[\tilde\chi(\tilde q, \tilde Y; \tilde T)]$
shifts to lower values of $\tilde Y$ when $\tilde q$ is risen.

The dependence of the real part of the susceptibility on $\tilde T$
and $\tilde q$ can also be analyzed by looking at the sum rules it
satisfies, which are defined as the energy--weighted integrals
\begin{equation}
\tilde m_n(\tilde q; \tilde T) = \int_0^\infty d\tilde\omega \,
\tilde\omega^n Re[\tilde\chi(\tilde q, \tilde\omega; \tilde T)] \ .
\label{j1}
\end{equation}

Sum rules could also be defined extending the integration range to
$\tilde\omega\in(-\infty,\infty)$, but then all the odd order ones
would trivially vanish because $Re[\tilde\chi(\tilde q, \tilde\omega;
\tilde T)]$ is an even function of $\tilde\omega$. With the previous
definition, and taking into account that the real part of the dynamic
susceptibility decreases as $2 \tilde q^2/\tilde \omega^2$ at large
energies, one readily notices that no sum rule with $n>0$
exists. However, an expansion around $\tilde\omega^{-1}=0$ displays
a next--to--leading term of order $\tilde\omega^{-4}$, so one can
still find a second energy weighted sum rule by subtracting the
leading contribution at high energies. Proceeding in this way one
finds the following relations
\begin{eqnarray}
\tilde m_0(\tilde q; \tilde T) & = & 
\int_0^\infty d\tilde\omega\, Re[\tilde\chi(\tilde q, \tilde\omega;
\tilde T)] = 0 \ ,
\label{j2} \\
\tilde m_2(\tilde q; \tilde T) & = & 
\int_0^\infty d\tilde\omega \,\tilde\omega^2 \left( 
Re[\chi(\tilde q,\tilde\omega; \tilde T)] - 
{2\tilde q^2 \over \tilde\omega^2} \right) \,= 0 \ .
\label{j3} 
\end{eqnarray}

These results actually apply to the total real part of the dynamic
susceptibility, including the condensate contribution. However, they
are general and therefore hold also for the $\tilde T=0$
susceptibility of the free Bose gas also. The condensate contribution
at finite temperature is equal to $n_0(\tilde T) Re[\tilde\chi(\tilde
q, \tilde\omega; \tilde T=0)]$, and thus its $\tilde m_0(\tilde q;
\tilde T)$ and $\tilde m_2(\tilde q; \tilde T)$ also vanish. In this
way, therefore, the above sum rules apply separately to the condensate
and non--condensate parts of the dynamic susceptibility, respectively.
Additionally, the fact that both sum rules vanish indicates that
$Re[\tilde\chi(\tilde q, \tilde\omega; \tilde T)]$ must have different
regions of positive and negative values, a feature that is in fact
general to all systems since the previous sum rules apply to both
correlated and uncorrelated systems at $\tilde T=0$ and finite
temperature.

Although the previous results hold for any system conserving the total
number of particles, it is difficult to obtain generic expressions for
higher order sum rules since then additional terms in the
$\tilde\omega^{-1}$ expansion of $Re[\tilde\chi(\tilde q,
\tilde\omega; \tilde T)]$ must be subtracted, and the coefficients of
this expansion are related to sum rules of the dynamic structure
function $\tilde S(\tilde q, \tilde\omega; \tilde T)$, which are
specific to each system. Furthermore, the simple subtraction of
$2\tilde q^2/\tilde\omega^2$ does not allow for the derivation of a
$\tilde m_1(\tilde q; \tilde T)$ sum rule, a problem that remains even
when higher orders in the $\tilde\omega^{-1}$ series expansion of
$Re[\tilde\chi(\tilde q, \tilde\omega; \tilde T)]$ are subtracted.

An alternative set of sum rules, valid for correlated Bose systems at
$\tilde T=0$ and $\tilde T>0$ and for the noninteracting Bose gas at
$\tilde T>0$, can be derived by subtracting to $\tilde\chi(\tilde q,
\tilde\omega; \tilde T)$ the $\tilde T=0$ dynamic susceptibility
$\tilde\chi^0(\tilde q, \tilde\omega; 0)$ of the free Bose
gas. As long as the real part is concerned, these become
\begin{equation}
\tilde M_n(\tilde q, \tilde T) = \int_0^\infty
d\tilde\omega\, \tilde\omega^n \left( 
Re[\tilde\chi(\tilde q, \tilde\omega; \tilde T)] - 
Re[\tilde\chi^0(\tilde q, \tilde\omega; 0)] \right) \ .
\label{k1}
\end{equation}

Using the explicit representation of the dynamic susceptibility in
terms of the dynamic structure function, the first $\tilde M_n(\tilde
q, \tilde T)$ moments can be easily derived and yield
\begin{eqnarray}
\tilde M_0(\tilde q; \tilde T) & = & 0 \ ,
\label{k2} \\
\tilde M_1(\tilde q; \tilde T) & = & 
-\int_{-\infty}^\infty d\tilde\omega \, 
\tilde S(\tilde q, \tilde\omega; \tilde T)
\ln\left( {\tilde\omega^2 \over \tilde q^4} \right) \ ,
\label{k3} \\
\tilde M_2(\tilde q; \tilde T) & = & 0 \ ,
\label{k4}
\end{eqnarray}

Interestingly enough, the expression of $\tilde M_1(\tilde q; \tilde
T)$ indicates that even though it is not know {\em a priori} whether
the dynamic structure function $\tilde S(\tilde q, \tilde\omega;
\tilde T)$ of a given system does have sum rules to all
orders~\cite{Holas}, still some combinations of them may exist. Notice
also that this sum rule is trivially satisfied by the condensate term,
since the condensate contribution to $\tilde S(\tilde q, \tilde\omega;
\tilde T)$ is a sum of two delta peaks centered at
$\tilde\omega=\pm\tilde q^2$ where the logarithm cancels. Therefore
and for the free Bose gas, the non--trivial part of $\tilde M_1(\tilde
q; \tilde T)$ must be satisfied by the non--condensate term
alone. Moreover and from the same argument given above, it is also
apparent that the total set of sum rules must be separately satisfied
by the condensate and non--condensate contributions to
$Re[\tilde\chi(\tilde q, \tilde\omega; \tilde T)]$. Finally, the
zeroth and second order sum rules $\tilde M_0(\tilde q; \tilde T)$ and
$\tilde M_2(\tilde q; \tilde T)$ are a direct consequence of the
results shown in Eqs.~(\ref{j2}) and~(\ref{j3}).

\begin{figure}
\begin{center}
\includegraphics[height=11cm]{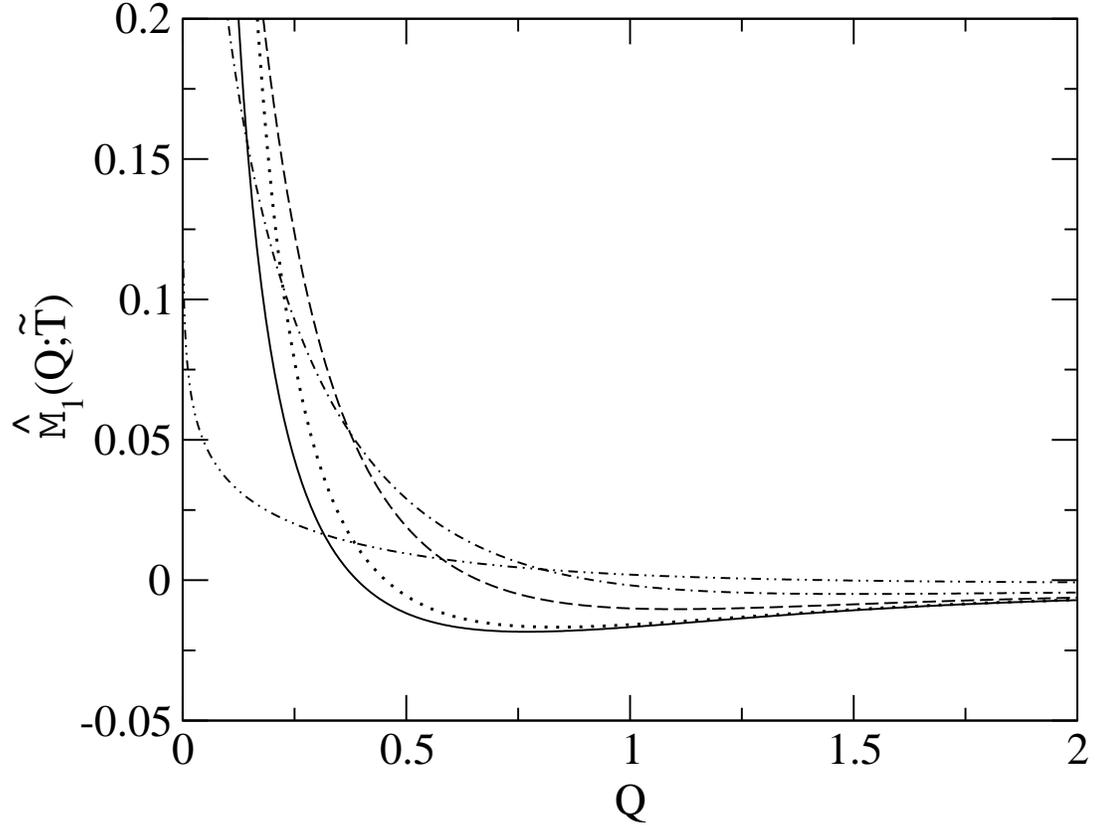}
\caption{First moment of the real part of the dynamic susceptibility
(Eq. (\ref{l1})) as a function of $Q$ and for several values of
temperature. The dotted, dashed, dot--dashed and dot--dot--dashed
lines stand for $\tilde T=7, 8, 10$ and $25$, respectively, while the
solid line corresponds to any $\tilde T$ below $\tilde T_c$.}
\label{fig-5}
\end{center}
\end{figure}

The fact that below $\tilde T_c$ the temperature dependence of both
the dynamic susceptibility and the response function can be extracted
as indicated in Eq.~(\ref{f2}) implies that an energy weighted sum
rule can also be defined in the form
\begin{equation}
\hat {\rm M}_1(Q; \tilde T) = {1 \over \tilde T^{3/2}} \tilde M_1(q;
\tilde T) \equiv \int_{-\infty}^\infty d\nu\, \tilde S(Q, \nu;
\tilde T) \ln\left( {\nu^2 \over Q^4} \right) \ ,
\label{l1}
\end{equation}
where $\tilde S(Q; \nu; \tilde T)=\tilde S_{nc}(\tilde q,\tilde\omega;
\tilde T)/\sqrt{\tilde T}$ is the non--condensate contribution to the
response of the free Bose gas written in the variables $Q$ and $\nu$.
Below the Bose--Einstein transition temperature, this factorization
removes any dependence on $\tilde T$ and thus $\hat{\rm M}_1$ becomes
a function of $Q$ alone. Above $\tilde T_c$, however, a dependence on
$\tilde T$ remains through the chemical potential $\tilde\mu(\tilde
T)$. $\hat{\rm M}_1(Q; \tilde T)$ is shown in Fig.~(\ref{fig-5}) for
several temperatures below and above $\tilde T_c$. In all cases, the
low $Q$ behaviour is divergent, a feature produced by the $Q^4$ term
inside the logarithm. On the other hand, the high $Q$ behaviour is
dominated by a slowly decreasing tail that approach $0$ from below. In
the intermediate region, $\hat{\rm M}_1(Q; \tilde T)$ presents a
minimum at some $Q$ that increases with the temperature, even though
it is hardly noticeable at high $\tilde T$'s.

In summary, it has been shown that the non--condensate contribution to
the real part of the dynamic susceptibility of the free Bose gas at
finite temperature can be evaluated in three different ways: by direct
integration, using a Kramers--Kronig relation and performing a Khanna
and Glyde -like series expansion. While the first two methods can be
used above and below the Bose--Einstein transition temperature, the
latter fails below $\tilde T_c$ since at those temperatures the
starting point in the series diverges. 

$Re[\tilde\chi(\tilde q, \tilde\omega; \tilde T\leq \tilde T_c)]$ is
shown to scale in two new variables $Q=\tilde q/\sqrt{\tilde T}$ and
$\nu=\tilde\omega/\tilde T$ from where its temperature dependence can
be extracted, a condition that is lost above $\tilde T_c$ due to the
finite value taken by the chemical potential $\tilde\mu(\tilde T)$.
Below $\tilde T_c$, the real part of the non--condensate contribution
to the dynamic susceptibility presents a discontinuity at
$\tilde\omega=\pm\tilde q^2$ produced by the singular behaviour of the
momentum distribution at $\tilde p\to 0$. When $\tilde T$ exceeds
$\tilde T_c$ this singularity in $n(\tilde p\to 0)$ is removed and as
a result the previous discontinuity is smeared out.  At high $\tilde
T>\tilde T_c$, $Re[\tilde\chi(\tilde q, \tilde\omega; \tilde T)]$
approaches the classical prediction computed from a Maxwell--Boltzmann
momentum distribution which for energies close to the quasielastic
peak $\tilde\omega=\tilde q^2$ can be brought to a form that scales in
a single variable $\tilde Y=(\tilde\omega/\tilde q-\tilde
q)/2\sqrt{\tilde T}$.  On the other hand, the leading contribution to
the high $\tilde q$ limit of $Re[\tilde\chi(\tilde q, \tilde\omega;
\tilde T)]$ can also be expressed in terms of $\tilde Y$ alone, a
property that is inherited from the large momentum transfer behaviour
of $\tilde S(\tilde q, \tilde\omega; \tilde T)$.

Low order energy weighted sum rules of the non--condensate
contribution to the real part of the dynamic susceptibility are also
derived and discussed. Since the long energy tails of
$Re[\tilde\chi(\tilde q, \tilde\omega; \tilde T)]$ decrease as
$1/\tilde\omega^2$, no sum rule of order higher than $0$
exists. However, a new set of sum rules, obtained by subtracting the
$\tilde T=0$ dynamic susceptibility of the free Bose gas, is proposed
and analyzed. These sum rules are general and apply to any system
conserving the total number of particles, and their analysis show that
in general the real part of the susceptibility must have different
regions where it changes sign. Despite the simplicity of the model
analyzed, the noninteracting Bose gas, some of the conclusions
drawn here are expected to enlighten some aspects of the finite
temperature dynamic susceptibility in weakly interacting systems,
which can be built starting from the non--interacting case in the
framework of the Random Phase Approximation.

\bigskip

\bigskip

\noindent{\bf Acknowledgments}

This work has been partially supported by the Austrian Science Fund
under grant No. P12832-TPH, DGICYT (Spain) grant No.  PB98-1247 and
the program SGR98-11 from {\em Generalitat de Catalunya}. One of us
wants to thank D. Mazzanti for giving a strong motivation to carry out
this work.  

\pagebreak

% %%%%%%%%%% THE BIBLIOGRAPHY MAY HELP, ALSO... %%%%%%%%%%

% %%%%%%%%%% HERE I PUT ALL THE FIGURES %%%%%%%%%%

\pagebreak

\pagebreak

\pagebreak

\pagebreak

\vfill\eject

\end{document}